\pdfoutput=1
\documentclass[letter,twocolumn,a4paper]{jpsj3}
\usepackage{bm}
\usepackage{graphics}
\usepackage{subfigure}

\title{Simple Real-Space Picture of Nodeless and Nodal s-wave Gap
Functions in Iron Pnictide Superconductors
}
\author{Toshikaze \textsc{Kariyado}\thanks{E-mail: kariyado@hosi.phys.s.u-tokyo.ac.jp} 
and Masao \textsc{Ogata}}
\inst{Department of Physics, University of Tokyo, 7-3-1 Hongo, Bunkyo-ku, Tokyo
113-0033\\
JST. TRIP, Sanbancho, Chiyoda-ku, Tokyo 102-0075}

\abst{We propose a simple way to parameterize the gap function in iron
pnictides. The key idea is to use orbital representation, not band
representation, and to assume real-space short-range pairing. Our
parameterization reproduces fairly well the structure of gap function
obtained in microscopic calculation. At the same time
the present parameterization is simple enough to obtain an intuitive
picture and to develop a phenomenological theory. We also discuss
simplification of the treatment of the superconducting state.
}

\kword{iron pnictides, superconductivity, gap function, multi-orbital
Hubbard model}

\begin{document}
\maketitle
What is new in the iron pnictide\cite{JPSJ.78.062001} as a high-T$_c$
superconducting
material is its multi-orbital nature. Several microscopic calculations
have been carried out to show that the five 3d orbitals of Fe atoms are
entangled\cite{Singh:2008,Kuroki:2009} and that the superconducting gap
function on the multiple
Fermi surfaces are very
complicated\cite{Yanagi:2008b,Ikeda:2008,Fuseya:2009,Graser:2009,Nomura:2008b,maier:224510}. Therefore,
it is desirable to construct a description of the gap function, which is
simple enough to obtain an intuitive picture for developing a
phenomenological theory, but also which is powerful enough to capture
its complicated multi-orbital nature. In fact, the gap functions of
several kinds of iron-pnictide superconductors are
not universal, i.e, nodeless in some materials and nodal in other
materials\cite{Hashimoto:2009,Nakai:2009,Hiraishi:2008,JPSJ.78.083712,JPSJ.78.103702,Ding:2008c,Kondo:2008,Fletcher:2008,Yamashita:2009}. Furthermore, there are
some controversial results on the existence of inter-band sign
reversal of
the gap function. For instance, although quasi-particle interference
measurements imply the existence of the inter-band sign
reversal\cite{Hanaguri:2009}, robustness of
T$_c$ against impurity
scattering suggests that there is no sign change between
bands\cite{Sato:2009,Onari:2009}. Having these experimental situations
in mind, we aim to construct a simple description of the gap function
that enables us to discuss such complexity in a simple manner.
We use orbital representation instead of band representation, and
assume real-space short-range pairing. Our parameterization reproduces
very well the structure of gap function obtained in microscopic RPA
calculation. We also discuss simplification of the gap function for
studying the superconducting state of iron pnictides.

We analyze the structure of the gap function based on 
the multi-orbital Hubbard model proposed by Kuroki {\it et
al.}\cite{Kuroki:2009} that is
downfolded from first-principle calculation. In this downfolding
scheme,
five bands around the Fermi energy are kept. The obtained five basis
wave
functions have the symmetry of Fe-3d orbitals, i.e.,
$d_{3z^2-r^2}$, $d_{zx}$,
$d_{yz}$, $d_{x^2-y^2}$ and $d_{xy}$. One thing to be noted here is
that the extended Brillouin zone is used, i.e, only one Fe
atom is contained in a unit cell. We use the Hamiltonian
$\mathcal{H}=\mathcal{H}_0+\mathcal{H}_I$ with
\begin{equation}
  \mathcal{H}_0=\sum_{\bm{i},\bm{j},\sigma,ab}
  t_{a,b;\bm{i},\bm{j}}c^{\dagger}_{\bm{i}a\sigma}c_{\bm{j}b\sigma}
  =
  \sum_{\bm{k}\sigma,ab}
  \varepsilon_{ab}(\bm{k})
  c^\dagger_{\bm{k}a\sigma}c_{\bm{k}b\sigma},
  \label{H0}
\end{equation}
where $t_{a,b;\bm{i},\bm{j}}$ is that obtained in the downfolding and
$\mathcal{H}_I$ is the standard onsite multi-orbital interaction with
$(U, U', J_H, J)$\cite{PhysRevLett.94.147005,Yanase:2005}. In
eq.~(\ref{H0}), indices $a$ and $b$ run through 0
to 4, which correspond to $d_{3z^2-r^2}$, $d_{zx}$, $d_{yz}$,
$d_{x^2-y^2}$ and $d_{xy}$ orbitals respectively. Details of the hopping
integrals are
different from system to system. In this paper, we mainly use the model
with hopping integrals shown in Table I of Kuroki {\it et
al.}\cite{Kuroki:2008}, in which the hopping integrals up to the
fifth-nearest neighbors are kept and the three dimensionality is
neglected. We call this model as typical-1111 model in the
following\cite{footnote1}. Full models for LaFeAsO, LaFePO and NdFeAsO
are also used.

Before analyzing the gap functions, we discuss the orbital- and
band-representations in the multi-orbital Hubbard model. The
Hamiltonian, eq.~(\ref{H0}), is written in the orbital representation
and $\mathcal{H}_0$ can be diagonalized by a momentum-dependent unitary
transformation. We call the transformed representation as band
representation, in which the free Green's function,
$\hat{G}_0(k)$ (5$\times$5 matrix form, $k=(i\omega_n,\bm{k})$), is
diagonalized. However, since the momentum dependence
of the unitary transformation is very severe in iron pnictides, the
orbital character strongly depends on the Fermi surface positions. This
causes the complexity when we use the band representation. In
contrast, we find that the orbital representation is often suitable to
discuss physics in real-space view. Although the Green's function,
$\hat{G}_0(k)$, in this representation has off-diagonal components
connecting different orbitals, it does not make serious problems when we
use the matrix form of Dyson-Gor'kov equation,
\begin{equation}
 \begin{pmatrix}
  (\hat{G}^0(k))^{-1}&\hat{\Delta}_{\bm{k}}\\
  \hat{\Delta}^\dagger_{\bm{k}}&-({}^t\hat{G}^0(-k))^{-1}
 \end{pmatrix}
 \begin{pmatrix}
  \hat{G}(k)&\hat{F}(k)\\
  \hat{\bar{F}}(k)&-{}^t\hat{G}(-k)
 \end{pmatrix}
 =\hat{1},
 \label{Dyson-Gorkov}
\end{equation} 
which is in the present case, 10$\times$10 matrix. Note that we only
consider the singlet channel and neglect the normal self-energy and
frequency dependence of the gap function in eq.~(\ref{Dyson-Gorkov}).

\begin{figure}[h]
 \begin{center}
  \includegraphics[height=50pt]{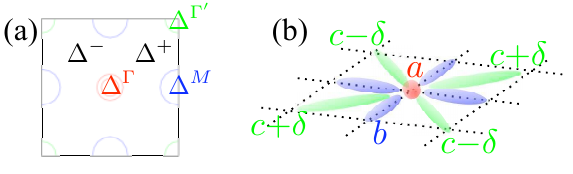}
  \caption{(Color online) (a) Schematic picture of the definition of
 $\Delta^\Gamma_i$, $\Delta^M_i$, $\Delta^{\Gamma'}_i$ and
  $\Delta_i^\pm$. (b) Real
  space picture of the $\Delta_{ii}(\bm{k})$.}
\label{fig1}
 \end{center}
\end{figure}

Now we begin to investigate the structure of $\hat{\Delta}_{\bm{k}}$.
The key idea is to use the orbital representation and to consider in
both real and momentum space. From now on, $\hat{\Delta}_{\bm{k}}$ is
always in the orbital representation. 
First, we consider the diagonal elements of $\hat{\Delta}_{\bm{k}}$,
i.e., the intra-orbital pairing in momentum space. Limiting our
consideration to the s-wave channel, what is important is the gap value
at $(0,0)$,$(\pi,0)$ and $(\pi,\pi)$ in the extended Brillouin zone,
since the Fermi surfaces of iron pnictides are small and enclose those
points. We denote the gap values on the orbital $i$ at $(0,0)$,
$(\pi,0)$ and $(\pi,\pi)$ as $\Delta^\Gamma_i$, $\Delta^M_i$ and
$\Delta^{\Gamma'}_i$, respectively. For later use, we also define the
difference of the gap value at $(\pm \pi/2,\pi/2)$ as
$\Delta^+_i-\Delta^-_i$ (see Fig.~\ref{fig1}(a)). 

Now we move on to the real space picture. 
In order to make a simple description of the gap functions, we assume
short-range pairings in the orbital representation. 
Then, $\Delta_{ii}(\bm{k})$ becomes
\begin{eqnarray}
 \Delta_{ii}(\bm{k})=a_i+2b_i(\cos k_x+\cos k_y)+4c_i\cos k_x\cos k_y\nonumber\\
 +4\delta_i\sin k_x\sin k_y, \label{diagonal}
\end{eqnarray}
where we have introduced three parameters, $a_i$, $b_i$ and $c_i$
representing on-site ($a_i$), nearest-neighbor ($b_i$), and
next-nearest-neighbor ($c_i$) pairings. Anisotropy parameter $\delta_i$
is introduced for the $d_{zx}$ and $d_{yz}$ orbitals ($i=$1, 2), since
they do not have 4-fold symmetry, i.e., we have
$\delta_1=-\delta_2=\delta$ and
$\delta_0=\delta_3=\delta_4=0$. Meanings of these
parameters are described in Fig.~\ref{fig1}(b).
Relation between the parameters
$(\Delta_i^\Gamma,\Delta_i^M,\Delta_i^{\Gamma'},\Delta_i^\pm)$ and
$(a_i,b_i,c_i,\delta_i)$ can be simply written as
 \begin{eqnarray}
  a_i&=&(\Delta_i^\Gamma+2\Delta_i^M+\Delta_i^{\Gamma'})/4,\ 
   b_i=(\Delta_i^\Gamma-\Delta_i^{\Gamma'})/8,\nonumber\\
  c_i&=&(\Delta_i^\Gamma-2\Delta_i^M+\Delta_i^{\Gamma'})/16,\ 
   \delta_i=(\Delta_i^--\Delta_i^+)/8.
   \label{d2abc}
 \end{eqnarray}
Note that the symmetry of 3d orbitals give $a_1=a_2$, $b_1=b_2$, 
$c_1=c_2$. 

Next, we consider the off-diagonal elements of $\hat{\Delta}_{\bm{k}}$, i.e.,
inter-orbital pairings. 
\begin{figure}[h]
 \begin{center}
  \includegraphics[height=80pt]{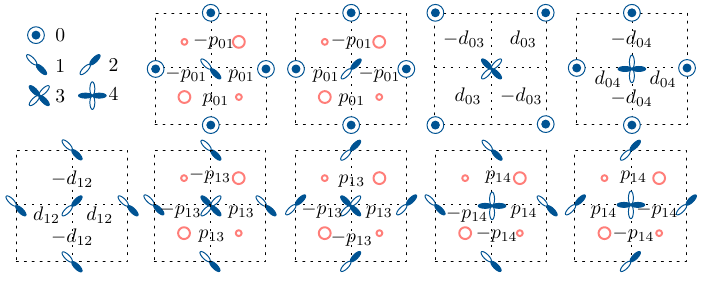}
  \caption{(Color online) Up left: Symmetries of the basis wave
  functions. Note that
  the coordinate of the original unit cell is used to describe the basis
  wave functions and they are rotated by $45^{\circ}$ from the usual
  definitions of the d-orbits on the square lattice. Others: Schematic
  real space pictures for the off-diagonal elements of
  $\hat{\Delta}(\bm{k})$. Red circle represents 
  the As sites.
  } 
  \label{fig2}
 \end{center}
\end{figure}
Assuming only the shortest possible pairing between the different
orbitals (Fig.~\ref{fig2}), we obtain
\begin{subequations}\begin{eqnarray}
  \Delta_{01}(\bm{k})&=&2ip_{01}(\sin k_x-\sin k_y),\\
  \Delta_{02}(\bm{k})&=&-2ip_{01}(\sin k_x+\sin k_y),\\
  \Delta_{03}(\bm{k})&=&2d_{03}(\cos (k_x+k_y)-\cos (k_x-k_y)),\\
  \Delta_{04}(\bm{k})&=&2d_{04}(\cos k_x-\cos k_y),\\
  \Delta_{12}(\bm{k})&=&2d_{12}(\cos k_x-\cos k_y),\\
  \Delta_{13}(\bm{k})&=&2ip_{13}(\sin k_x-\sin k_y),\\
  \Delta_{23}(\bm{k})&=&2ip_{13}(\sin k_x+\sin k_y),\\
  \Delta_{14}(\bm{k})&=&2ip_{14}(\sin k_x+\sin k_y),\\
  \Delta_{24}(\bm{k})&=&-2ip_{14}(\sin k_x-\sin k_y).
 \end{eqnarray} \label{swavedelta}\end{subequations}
Here the symmetries of orbitals are important for the sign of
pairings as shown in Fig.~\ref{fig2}. Furthermore, the
matrix elements concerning $d_{zx}$ and $d_{yz}$ orbitals ($i=$1, 2) are
determined by taking account of As atoms which exist above and below the
Fe plane, because the $d_{zx}$/$d_{yz}$ orbitals are odd in z-direction
and they do not have matrix elements with others without As. 
Characteristic feature found in eqs.~(\ref{swavedelta}) is that the
off-diagonal
elements have p-wave or d-wave symmetry. This is a direct consequence
of the symmetry of the basis wave functions and the fact that the
diagonal elements have s-wave symmetry. 
In the case where the
diagonal elements have d-wave symmetry, we can do the similar analysis.

In order to determine the parameters in eqs.(\ref{diagonal}) and
(\ref{swavedelta}), we calculate the gap
function $\hat{\Delta}^{\mathrm{RPA}}_{\bm{k}}$, 
by solving linearized Eliashberg
equation with the effective interaction obtained within RPA. In
this calculation, we follow the formalisms in Kuroki {\it et
al.}\cite{Kuroki:2009}. We use
the filling $n=6.1$, temperature $T=0.02$ eV and Coulomb
parameters $(U,U',J_H,J)=(1.0,0.6,0.2,0.2)$ eV. The Brillouin
zone is divided into 32$\times$32 meshes and 512 Matsubara frequencies are
used. Then, using the gap function
$\hat{\Delta}^{\mathrm{RPA}}_{\bm{k}}$, we obtain 
$(\Delta_i^\Gamma,\Delta_i^M,\Delta_i^{\Gamma'},\Delta_i^\pm)$ from
which the parameters $(a_i,b_i,c_i,\delta_i)$ are determined through
eqs.~(\ref{d2abc}). Parameters $p_{ij}$ and $d_{ij}$ are determined from
$\Delta_{ij}^{\mathrm{RPA}}(\bm{k})$ ($i\neq j$) similarly.
The obtained results for a typical-1111 model are
shown in Table.~\ref{params_table}.

\begin{figure}[h]
 \begin{center}
  \includegraphics[height=130pt]{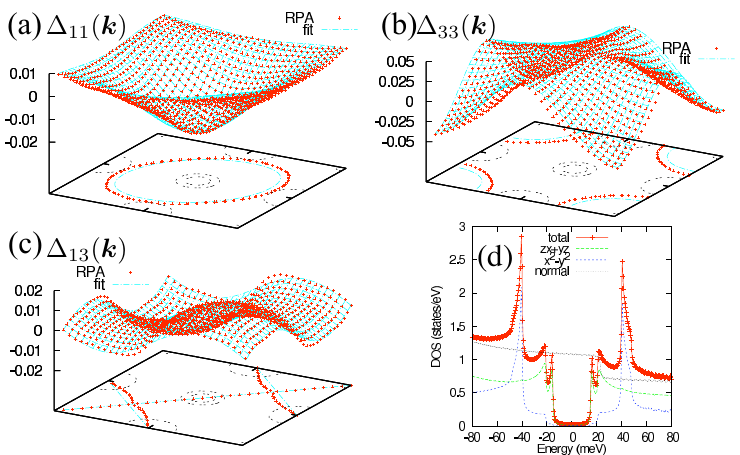}
  \caption{(Color online) (a-c) Calculated $\Delta_{11}(\bm{k})$,
  $\Delta_{33}(\bm{k})$
  and $\Delta_{13}(\bm{k})$ in RPA and fitted ones for typical-1111
  model. (d) DOS for typical-1111 model calculated with 2048$\times$2048
  k-space meshes.}
  \label{fig3}
 \end{center}
\end{figure}

In Fig.~\ref{fig3}, we compare the present simple description
$\Delta_{11}(\bm{k})$, $\Delta_{33}(\bm{k})$ and $\Delta_{13}(\bm{k})$
with those in $\hat{\Delta}_{\bm{k}}^{\mathrm{RPA}}$ determined
microscopically.
We can see that our fitting is in good agreement with
$\hat{\Delta}^{RPA}_{\bm{k}}$. 
Although some of the off-diagonal elements in $\hat{\Delta}(\bm{k})$ do
not show a good coincidence, their magnitudes are small so that it 
does not lead to a serious problem. It is rather surprising
that the gap function can be expressed only with the small number of
parameters representing the real-space pairing up to
next-nearest-neighbor. In contrast, when the gap functions are
re-expressed in the band representation, they become very complicated for
getting an intuitive picture.
\begin{table*}[htb]
 \begin{center}
  \caption{Parameters obtained in RPA calculation. Typical 6.1, typical
  6.3 are the results for typical-1111 model with the filling $n=$6.1 and
  6.3 respectively. The result for LaFePO is obtained with the filling
  $n=$6.1.}\label{params_table}
  \begin{tabular}{r|c|c|c|c|c|c|c|c|c}
   \hline
   & $a_1$ & $b_1$ & $c_1$ &$\delta$ & $a_3$ & $b_3$ & $c_3$ & $d_{12}$ &
   $p_{13}$\\
   \hline
   typical 6.1 & 0.000746 & -0.00382 & -0.00155 & 0.000751 & 0.0213 &
			   0.0111 & -0.00474 & -0.000536 & 0.00446 \\
   \hline
   typical 6.3 & 0.00185& -0.000598 & -0.00517 &
	       0.000895 & -0.00541& 0.0117 & 0.000315
			   & -0.000248 & 0.00642 \\
   \hline
   LaFePO &0.00596 & -0.00301 & -0.00107 & 0.000504 & -0.00716 & 0.00235 &
			       -0.000441 & -0.00220 & 0.00166 \\
   \hline
  \end{tabular}
 \end{center}
\end{table*}
In Fig.~\ref{fig3}(d), we also plot the density of states (DOS)
calculated with the fitted gap functions. The result shows nodeless 
two-gap behavior where the size of the smaller gap is about half of the
larger gap.

We have checked the present parameterization for various cases to find
that the fitting is very good for most of the cases. Exception is the
case when we have too strong spin fluctuation on $d_{x^2-y^2}$
orbitals, for example in the model for NdFeAsO. In this case,
$\Delta_{33}(k)$ has strong momentum dependence
and eq.~(\ref{diagonal}) is not enough to reproduce its strong momentum
dependence. However, it is well known that the RPA estimation of the
effective interaction gives stronger momentum dependence of
$\hat{\Delta}(\bm{k})$ compared with other approximations. Therefore, we
expect that the actual momentum dependence of
$\Delta_{33}(\bm{k})$ in this case is much
milder and the present parameterization works as well. In
addition, the damping effect caused by impurities and the
strong-correlation effects neglected in RPA generally give smoother gap
functions. 

In the following, we consider the possible simplification of the gap
functions. 
For the typical-1111 model and the models for LaFeAsO and
NdFeAsO\cite{footnote1}, we expect that only $d_{zx}$/$d_{yz}$ and
$d_{x^2-y^2}$
orbitals are necessary since they contribute most of the DOS near the
Fermi energy. In this case, 
we need only parameters ($a_1$,
$b_1$, $c_1$, $\delta$, $a_3$, $b_3$, $c_3$, $d_{12}$, $p_{13}$). 
We find that DOS calculated with keeping these parameters and
setting others to zero can reproduce the DOS in
Fig.~\ref{fig3}(d). Furthermore, even if we assume
$d_{12}=p_{12}=0$, the obtained DOS reproduces most of the features in
Fig.~\ref{fig3}(d), although there is a slightly different structure
for the smaller gap.
Since the
resultant gap function is so simple, it is very useful to obtain an
intuitive picture.
Note that the above simplification can not be applied when $d_{3z^2-r^2}$ or
$d_{xy}$ contribute to the DOS around Fermi energy, for example, in the
case of LaFePO.

\begin{figure}[h]
 \begin{center}
  \includegraphics[height=70pt]{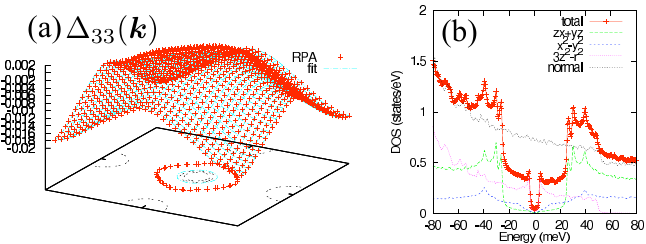}
  \caption{(Color online) (a) Calculated $\Delta_{33}(\bm{k})$ and
  fitted one for LaFePO
  model. (b) DOS for LaFePO model calculated with
  512$\times$512$\times$32 k-space meshes.}
  \label{fig4}
 \end{center}
\end{figure}

Next, we study the the difference between the nodal and nodeless s-wave gap
in terms of the present parameterization. The existence of the nodal
s-wave is
theoretically predicted for the model for
LaFePO\cite{Kuroki:2009}. Thus, we carried
out similar calculation using the model of LaFePO. Here we use
three-dimensional model and divide the Brillouin zone into
32$\times$32$\times$4 meshes. We find that the most
significant difference from the typical-1111 model appears
in $\Delta_{33}(\bm{k})$, which is shown in Fig~\ref{fig4} (see
also Table.~\ref{params_table}). 
Comparing Figs.~\ref{fig4} and \ref{fig3}, we can see that
$\Delta_{33}(\bm{k})$ for LaFePO model has weaker momentum dependence
than that for the typical-1111 model. In terms of our parameterization,
$c_3$ is
smaller in LaFePO. Due to this difference, $\Delta_{33}(\bm{k})$ at
$(\pi,0)$ is positive in
the typical-1111 model and negative in LaFePO. Note that
$\Delta_{11}(\bm{k})$ at $(\pi,0)$ is positive for both models. 
Figure~\ref{fig4}(b) shows the obtained DOS for LaFePO. We can see
that (1) the partial DOS for the $d_{zx}$/$d_{yz}$ and $d_{x^2-y^2}$
orbitals show nodal behavior and (2) there is
large remaining DOS in the gap around 20 meV since only a very small gap
($\lesssim$ 2 meV) appears in the $d_{3z^2-r^2}$ orbital.

\begin{figure}[h]
 \begin{center}
  \includegraphics[height=70pt]{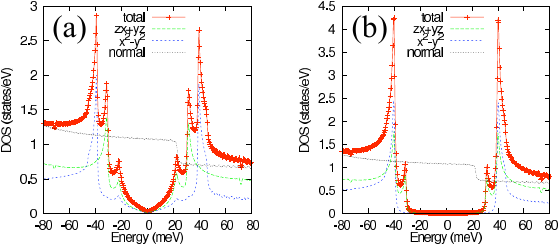}
  \caption{(Color online) DOS for the hypothetical gap function of
  $(a_1,a_3)=(\Delta,\Delta)$ (a) and $(\Delta,-\Delta)$ (b) calculated
  with 2048$\times$2048 k-space meshes and $\Delta=$40 meV.}
  \label{fig5}
 \end{center}
\end{figure}

To obtain a clear view on how the node evolves, we calculate DOS
with a hypothetical gap function, with only
$(a_1,a_3)=(\Delta,-\Delta)$. This means
that we consider the totally momentum-independent s-wave gap in the
orbital
representation but with different sign on $d_{zx}/d_{yz}$
orbitals and $d_{x^2-y^2}$ orbitals. Interestingly, obtained DOS shows clear
nodal behavior (Fig.~\ref{fig5}(a)) although we input the gap function
without any momentum dependence. This is because the orbital character
varies along the positions of the
Fermi surface. Some part of the Fermi surface enclosing $(\pi,0)$ has
$d_{zx}/d_{yz}$
character and the gap function is $+\Delta$, while other part of that
Fermi surface has $d_{x^2-y^2}$ character and the gap function is
$-\Delta$. In this case, nodes appear on the Fermi surface. Although the
gap with $(a_1,a_3)=(\Delta,-\Delta)$ is not the result of microscopic
calculation, this argument captures the nature of
the node obtained for LaFePO model. For comparison, we show the
result with $(a_1,a_3)=(\Delta,\Delta)$ in Fig.~\ref{fig5}(b). It shows
clear full-gap feature and no nodal behavior as expected. 

\begin{figure}[h]
 \begin{center}
  \includegraphics[height=120pt]{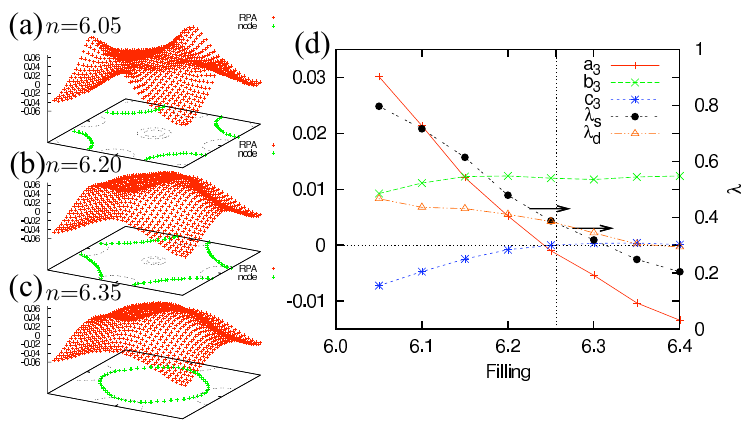}
  \caption{(Color online) (a-c) Doping dependence of
  $\Delta_{33}(\bm{k})$. (d) Doping
  dependence of $a_3$, $b_3$ and $c_3$. Eigenvalues of the linearized
  Eliashberg equation for s-wave pairing channel $\lambda_s$ and for
  d-wave pairing channel $\lambda_d$ are also plotted.}
  \label{fig6}
 \end{center}
\end{figure}

Finally we investigate the doping dependence of the gap
function within RPA calculation concentrating on the electron-doping
side. We use the typical-1111 model and treat
the doping as the rigid band shift. As we can see from
Fig.~\ref{fig6}, $\Delta_{33}(\bm{k})$ becomes flatter and flatter
with doping. Figure~\ref{fig6}(d) shows $a_3$, $b_3$ and $c_3$
in $\Delta_{33}(\bm{k})$ as a function of doping. 
The most important behavior
is that of $c_3$. With doping the absolute value of
$c_3$ becomes smaller and $c_3$ disappears around 
$n\sim 6.25$. Actually, the strongest superconducting instability channel
changes from s-wave to d-wave around $n\sim 6.25$. (See the eigenvalues
of the linearized Eliashberg equation for each channel $\lambda_s$ and
$\lambda_d$ plotted in Fig.~\ref{fig6}(d).) This suggests
that the
next-nearest-neighbor pairing of $d_{x^2-y^2}$ orbital plays important
role in stabilizing s-wave channel. This is a good example to show the
convenience of our parameterization in obtaining an intuitive picture
for the gap function.

In summary we have investigated the structure of the gap function in
iron pnictides. We have shown that the combination of the orbital
representation and real space view leads to a powerful and convenient
expression for the gap functions. This expression is powerful enough to
reproduce the results of microscopic calculation and to discuss the
variety of the gap functions. At the same time it is simple enough to
make an intuitive picture and to calculate physical quantities in the
superconducting state through eq.~(\ref{Dyson-Gorkov}). 
Calculation of the physical quantities in this
formalism should give interesting information on whether the gap
function is s-wave or d-wave, or the gap function is s$_{++}$ or
s$_{+-}$. Similar analysis on 122- and 11- system will also give
important information.

\section*{Acknowledgment}
We thank S.~Onari and Y.~Yanase for useful comments and discussions. We
also thank K.~Kuroki for useful comments and sharing the data of the
downfolded models.



\begin{thebibliography}{10}

\bibitem{JPSJ.78.062001}
K. Ishida, Y. Nakai, and H. Hosono:  J. Phys. Soc. Jpn. {\bf 78} (2009) 062001.
\bibitem{Singh:2008}
D. Singh and M. Du:  Phys. Rev. Lett. {\bf 100} (2008) 237003.
\bibitem{Kuroki:2009}
K. Kuroki, H. Usui, S. Onari, R. Arita, and H. Aoki:  Phys. Rev. B {\bf 79}
  (2009) 224511.
\bibitem{Yanagi:2008b}
Y. Yanagi, Y. Yamakawa, and Y. Ono:  J. Phys. Soc. Jpn {\bf 77} (2008) 123701.
\bibitem{Ikeda:2008}
H. Ikeda:  J. Phys. Soc. Jpn. {\bf 77} (2008) 123707.
\bibitem{Fuseya:2009}
Y. Fuseya, T. Kariyado, and M. Ogata:  J. Phys. Soc. Jpn {\bf 78} (2009)
  023703.
\bibitem{Graser:2009}
S. Graser, T.~A. Maier, P.~J. Hirschfeld, and D.~J. Scalapino:  New J. Phys.
  {\bf 11} (2009) 025016.
\bibitem{Nomura:2008b}
T. Nomura:  J. Phys. Soc. Jpn. {\bf 78} (2009) 034716.
\bibitem{maier:224510}
T.~A. Maier, S. Graser, D.~J. Scalapino, and P.~J. Hirschfeld:  Phys. Rev. B
  {\bf 79} (2009) 224510.
\bibitem{Hashimoto:2009}
K. Hashimoto, M. Yamashita, S. Kasahara, Y. Senshu, N. Nakata, S. Tonegawa, K.
  Ikada, A. Serafin, A. Carrington, T. Terashima, H. Ikeda, T. Shibauchi, and
  Y. Matsuda:  arXiv:0907.4399.
\bibitem{Nakai:2009}
Y. Nakai, T. Iye, S. Kitagawa, K. Ishida, S. Kasahara, T. Shibauchi, Y.
  Matsuda, and T. Terashima:  arXiv:0908.0625.
\bibitem{Hiraishi:2008}
M. Hiraishi, R. Kadono, S. Takeshita, M. Miyazaki, A. Koda, H. Okabe, and J.
  Akimitsu:  J. Phys. Soc. Jpn. (2009) 023710.
\bibitem{JPSJ.78.083712}
H. Fukazawa, Y. Yamada, K. Kondo, T. Saito, Y. Kohori, K. Kuga, Y. Matsumoto,
  S. Nakatsuji, H. Kito, P.~M. Shirage, K. Kihou, N. Takeshita, C.-H. Lee, A.
  Iyo, and H. Eisaki:  J. Phys. Soc. Jpn. {\bf 78} (2009) 083712.
\bibitem{JPSJ.78.103702}
M. Yashima, H. Nishimura, H. Mukuda, Y. Kitaoka, K. Miyazawa, P.~M. Shirage, K.
  Kihou, H. Kito, H. Eisaki, and A. Iyo:  J. Phys. Soc. Jpn. {\bf 78} (2009)
  103702.
\bibitem{Ding:2008c}
H. Ding, P. Richard, K. Nakayama, T. Sugawara, T. Arakane, Y. Sekiba, A.
  Takayama, S. Souma, T. Sato, T. Takahashi, Z. Wang, X. Dai, Z. Fang, G.~F.
  Chen, J.~L. Luo, and N.~L. Wang:  Europhys. Lett. {\bf 83} (2008) 47001.
\bibitem{Kondo:2008}
T. Kondo, A.~F. Santander-Syro, O. Copie, C. Liu, M.~E. Tillman, E.~D. Mun, J.
  Schmalian, S.~L. Bud'ko, M.~A. Tanatar, P.~C. Canfield, and A. Kaminski:
  Phys. Rev. Lett. {\bf 101} (2008) 147003.
\bibitem{Fletcher:2008}
J. Fletcher, A. Serafin, L. Malone, J. Analytis, J. Chu, A. Erickson, I.
  Fisher, and A. Carrington:  Phys. Rev. Lett. {\bf 102} (2009) 147001.
\bibitem{Yamashita:2009}
M. Yamashita, N. Nakata, Y. Senshu, S. Tonegawa, K. Ikada, K. Hashimoto, H.
  Sugawara, T. Shibauchi, and Y. Matsuda:  arXiv:0906.0622 (2009) .
\bibitem{Hanaguri:2009}
T. Hanaguri:  private communication.
\bibitem{Sato:2009}
M. Sato, Y. Kobayashi, S.~C. Lee, H. Takahashi, and Y. Miura:  arXiv:0907.3007.
\bibitem{Onari:2009}
S. Onari and H. Kontani:  arXiv:0906.2269.
\bibitem{PhysRevLett.94.147005}
M. Mochizuki, Y. Yanase, and M. Ogata:  Phys. Rev. Lett. {\bf 94} (2005)
  147005.
\bibitem{Yanase:2005}
Y. Yanase, M. Mochizuki, and M. Ogata:  J. Phys. Soc. Jpn. {\bf 74} (2005) 430.
\bibitem{Kuroki:2008}
K. Kuroki, S. Onari, R. Arita, H. Usui, Y. Tanaka, H. Kontani, and H. Aoki:
  Phys. Rev. Lett. {\bf 101} (2008) 087004.
\bibitem{footnote1}
  Actually, this model gives a dispersion relation that is between LaFeAsO and
  NdFeAsO. Therefore, this "typical--1111" model should be distinguished from
  the full model for LaFeAsO and NdFeAsO.
\end{thebibliography}
\end{document}